# NEWS & VIEWS

## COSMOLOGY

# The Universe's skeleton sketched

Eric V. Linder

**The deepest and clearest maps yet of the Universe's skeleton of dark-matter structure present a picture broadly in concord with favoured models — although puzzling discrepancies remain.**

Astronomers use every resource at their disposal to construct an image of the Universe: from the microwave radiation of the cooling Big Bang, past the visible wavelengths of light seen by the casual viewer of the night sky, to the γ-rays of the most powerful explosions of collapsing stars and ravenous black holes. They don't even have to see an object directly to detect its presence. In a manner akin to the Polynesian seafarers who sense islands out of their sight through the deflected direction of ocean waves, cosmologists can map a concentration of the Universe's unseen mass through the gravitational deflection of light coming from sources behind it. On *Nature*'s website today[1], Massey *et al.* publish the latest and most detailed atlas of these 'dark-matter' congregations.

Dark matter is thought to comprise as-yet undiscovered elementary particles, and is a central prediction of the current 'concordance' model of the Universe's make-up. Here, it acts as a skeleton around which bright matter — galaxies and clusters of galaxies — assembles. There is abundant indirect evidence for structures made of dark matter. The gravitational deflection, or lensing, technique has already been used to make maps of dark matter around individual galaxies and clusters, most dramatically by Clowe *et al.*[2]. Low-resolution reconstructions of the skeleton of a larger chunk of the nearby Universe have also been carried out using ground-based telescopes.

Massey *et al.*[1] present an analysis of gravitational-lensing data from the Cosmic Evolution Survey, COSMOS, that were acquired using the Hubble Space Telescope. These data cover an area of sky about eight times the size of the full Moon, and represent the first wide-sky, space-based survey of dark matter. This is a great step forward, as observations from space avoid the distortions and time variations that Earth's atmosphere imposes on the astronomical signal. (The effect of these perturbations is rather as though our seafarer friend had to contend with a nearby canoe's paddlers roiling the ocean when studying his waves.) The new maps consequently have a much higher resolution than the best ground-based observations, with four times the density of sources than in previous studies. They also go deeper into the Universe, looking back to a time when the Universe was about half the age it is now, at a redshift $z=1$. (The redshift is a measure of the expansion of the Universe; the higher the redshift of a cosmological object, the smaller and younger the Universe was when the object sent out its signal.)

To assess the third dimension of their map, the depth of the field of view, astronomers cannot move their canoe and use triangulation. They rely instead on mathematics to derive the distance of the gravitational lens by means of its focal length — which is known from Einstein's general theory of relativity — and the distance to the background source galaxy whose light is probing the dark matter. Massey *et al.* combine the COSMOS observations with follow-up observations from ground-based telescopes in 15 wavelength bands, and so are able to estimate redshifts of the galaxy sources all the way back to $z=3$, when the Universe was just a sixth of its present age.

With this information, the pattern of dark-matter concentrations can be mapped crudely in three dimensions by slicing the data into three redshift shells. Although this is just a first step in 'cosmic tomography', the advantage of using a space telescope is that the source density in each slice of this three-dimensional map is nearly as high as for the 'unsliced' maps — two-dimensional projections — expected from the next generation of ground-based surveys.

Massey and colleagues' three-dimensional map can be studied on its own merits to provide information about the pattern and evolution of dark-matter structures, or can be compared with observations of the bright matter through visible light and X-rays. The dark-matter distribution shows modest evidence for a connected, filamentary distribution — exactly the kind of skeleton predicted by concordance cosmology for stars and galaxies to assemble around.

A note of caution, however: it is still early days, and spurious noise affects the weaker features in the map (at the level of around 2.9 standard deviations). There are also striking examples of both beautiful agreement and puzzling disagreement between the dark and bright maps. Areas bright in X-rays, indicating the highest concentrations of gas heated by falling into the strongest gravitational fields, almost always overlap with the dark-matter concentrations — but not absolutely always. Conversely, dark-matter concentrations sometimes seem to have no corresponding bright matter. For the dark-matter skeleton of mass in the Universe, flesh sometimes occurs without supporting bones, and bones without surrounding flesh. That is perfectly possible. However, these data, together with forthcoming ground-based data from the lower-resolution, but wider-area Canada–France–Hawaii Telescope Legacy Survey[3], will need to be carefully compared with computational simulations of the Universe to analyse consistencies and discrepancies.

Cosmologists are getting a foretaste of a new ability to map the dark Universe, despite the fact that the gravitational-lens mapping technique is still showing some growing pains: even the Hubble Space Telescope observations required significant corrections, because of radiation damage to the detectors. Radiation-resistant detectors have already been developed for a proposed space mission, the Supernova/Acceleration Probe (SNAP)[4], that would extend COSMOS-type maps of the dark-matter skeleton to an area 2,500 times larger than that covered by Massey *et al.*[1]. These maps would also be both deeper and more finely resolved. But for now, the COSMOS data give astronomers an exciting new view of the dark Universe. ■

Eric V. Linder is at the University of California and Lawrence Berkeley National Laboratory, Berkeley, California 94720, USA.
e-mail: evlinder@lbl.gov